\documentclass[preprint,prd,showkeys,showpacs,superscriptaddress,longbibliography]{revtex4-1}
\pdfoutput=1
\usepackage{amsfonts,amsmath,amssymb}
\usepackage{enumerate}
\usepackage{hyperref}
\usepackage{graphicx}
\usepackage{caption}
\usepackage{subcaption}
\usepackage{bibentry}
\usepackage{color}
\newcommand{\be}{\begin{equation}}
\newcommand{\ee}{\end{equation}}

\newcommand{\bea}{\begin{eqnarray}}
\newcommand{\eea}{\end{eqnarray}}

\newcommand{\bal}{\begin{align}}
\newcommand{\eal}{\end{align}}

\newcommand{\bes}{\begin{subequations}}
\newcommand{\ees}{\end{subequations}}

\newcommand{\nn}{\nonumber}

\newcommand{\ra}{\rightarrow}

\begin{document}

\title{Time-evolution of holographic entanglement entropy \\and metric perturbations}
\author{Nakwoo Kim}
\email{nkim@khu.ac.kr}
\affiliation{Department of Physics
and Research Institute of Basic Science, \\ Kyung Hee University,
Seoul 02447, Korea }
\affiliation{School of Physics, Korea Institute for Advanced Study, Seoul 02455, Korea}
\author{Jung Hun Lee}
\affiliation{Department of Physics and Research Institute of Basic Science, \\ Kyung Hee University,
Seoul 02447, Korea }

\begin{abstract}
We study the holographic entanglement entropy under small deformations of AdS, including time-dependence.
It is found through perturbative analysis that the divergent terms are not affected and the change appears only
in the finite terms. We also consider the entanglement thermodynamic first law, and calculate the entanglement
temperature and confirm that it is inversely proportional to the size of the entangling region.
\end{abstract}

\maketitle

\section{Introduction}
The holographic principle \cite{Hooft:1993gx,Susskind:1994vu,Maldacena:1997re} relates a certain strongly-coupled field theory to a quantum gravity system
in one higher dimensions. It is an intriguing proposal, when we note  the possibility of spacetime 
emerging naturally from a quantum system without gravitational degrees of freedom {\it per se}. 
The holographic entanglement
entropy (HEE) prescription \cite{Ryu:2006bv} provides a versatile tool in this line of study, since unlike 
other probes like Wilson loops which are dual to fundamental strings, it involves 
model-independent and universal concepts like quantum entanglement and volume-minimizing submanifold etc.

Another valuable insight into this relation is motivated by the so-called AdS/MERA correspondence \cite{Swingle:2009bg}.
MERA, or multi-scale entanglement renormalization ansatz, is a particular version of tensor network method in the study 
of critical lattice system \cite{Vidal:2007hda}. In MERA, one considers the real-space renormalized lattice system together 
with the original lattice system and extremizes the energy of the entire system, from UV to IR. In this picture
a holographic direction in AdS/CFT language is incorporated by construction, and one can argue that the metric 
in the entire space is the hyperbolic space, which is just the constant time slice of AdS. It is certainly of great
interest to promote such observations to full-fledged time-dependent backgrounds. Namely, we hope to be
able to see propagating gravitons constructed out of time-dependent process in a strongly-coupled quantum
system. The present work is a modest step towards that goal and we will study the time-evolution of HEE in time-dependent
backgrounds.

The Ryu-Takayanagi proposal has been generalized to covariant formulae to suit 
generic backgrounds in \cite{Hubeny:2007xt}, and in this paper we will study the behavior of extremal 
codimension-two surfaces in backgrounds which are small deformations of pure AdS spacetime.
 There exist works which studied the evolution of HEE in time-dependent setups already, but as far
as we know they mostly use a rather special class of metric called AdS-Vaidya solutions, or make
use of numerically constructed solutions \cite{Balasubramanian:2010ce,Balasubramanian:2011ur,
Balasubramanian:2011at,Allais:2011ys,Keranen:2011xs,Caceres:2012em,Baron:2012fv,
Hubeny:2013hz,Liu:2013iza,Liu:2013qca,Hubeny:2013dea,Alishahiha:2014jxa,Bai:2014tla,
daSilva:2014zva,Ugajin:2014nca,Keranen:2015fqa,Ziogas:2015aja}. 
Conceptually our study is more similar to papers such as \cite{Leichenauer:2015nxa,Mishra:2015cpa}
which adopted a perturbative approach, but we are interested in time-dependendent backgrounds.

In Sec.2 we start by computing the HEE for cap-like regions in global coordinate of AdS, and sketch how
HEE can be computed perturbatively in backgrounds with time-dependence. For simplicity we restrict ourselves
to spherically-symmetric backgrounds. In Sec.3 we apply the method to AdS-Schwarzschild black holes and
also to the time-periodic solutions of AdS-scalar field system constructed in \cite{Maliborski:2013jca,Kim:2014ida}.
One generic feature we obtained is that the effect of metric perturbation on HEE starts with the finite part, 
and there is no change in the leading-order or logarithmically divergent terms. In Sec.4 we use our results to 
compute the holographic entanglement temperature and verify that it is inversely proportional to the size of 
the entangling region, with a universal coefficient. We conclude with a couple of comments in Sec.5.
\section{HEE with metric perturbations}
\subsection{HEE in pure AdS}
As a warm-up and also to set up the notation let us consider the holographic entanglement entropy of pure AdS in generic dimensions. The metric convention we take 
for $AdS_{d+1}$ spacetime with radius $\ell$ is
\bea
ds^2=\frac{\ell^2}{\cos^2x}(-dt^2+dx^2+\sin^2 x(d\theta^2+\sin^2\theta d\Omega_{d-2}^2)),
\label{ads}
\eea
where for later convenience we split the $S^{d-1}$ metric into polar angle $\theta$ and unit-radius sphere $S^{d-2}$ with metric $d\Omega_{d-2}^2$. 

In this paper we choose the entangling region on the boundary to be a cap-like one
defined by $0\le \theta <\theta_0$ and at 
boundary time $t=t_0$. 
The holographic entanglement entropy (HEE) is then the extremized area from the following functional of $x(\theta)$,
\bea
\text{Area}=\ell^{d-1}\text{vol}(S^{d-2})\int^{\theta_0}_0 d\theta\frac{( \sin x\sin\theta)^{d-2}}{\cos^{d-1}x}\sqrt{\left(\frac{dx}{d\theta}\right)^2+\sin^2 x} \,\, .   \label{area}
\eea
Without losing generality we assume $\theta_0<\pi/2$. The most general form
of the solution derived from the action \eqref{area} is difficult to find, 
but one can check the following function satisfies the Euler-Lagrange equation from 
\eqref{area} for any $d$ \cite{Bakas:2015opa}:
\bea
x(\theta)=\sin^{-1} \left(\frac{a}{\cos\theta}\right) . 
\label{cl}
\eea
This solution is related through a conformal transformation
to the hemisphere solutions of a 
spherical-shape entangling surface in Poincar\'e coordinates. We may call them 
{\it constant-latitude} solutions because $\sin x \cos\theta = a$. Since the boundary is
at $x=\pi/2$, we have $a=\cos\theta_0$. These curves are plotted in Fig.\ref{constlat}, using $x$ as the radial 
coordinate.

We will consider small deformations of $AdS$ background and see how the 
minimal area solutions \eqref{cl} changes accordingly. 
Instead of the original parametrization, we find that it is more advantageous to use
the following variables.
\bea
z :=\cos\theta, \quad\rho :=\sin x \, .
\eea
Then the area functional becomes
\bea
\text{Area}=\ell^{d-1}\text{vol}(S^{d-2})\int^1_{\cos^{-1}\theta_0} dz 
\frac{\rho^{d-2}(1-z^2)^{(d-2)/2}}{(1-\rho^2)^{d/2}}\sqrt{(\rho')^2+\frac{\rho^2(1-\rho^2)}{1-z^2}}.
\label{area2}
\eea
To calculate HEE we need to substitute the solution $\rho=a/z$ back into \eqref{area2}, and perform the integral.  As it is usually the case with AdS/CFT, the integral is divergent near the boundary $x=\pi/2$ and
 a regularization process is required. 
We introduce a cutoff at $x_m=\pi/2-\epsilon$. When applied to the reference solution
\eqref{cl},
\bea
\theta_{m}=\theta_0-\frac{1}{2}\epsilon^2\cot\theta_0 ,\quad \text{i.e.} \quad z_{m}=
a\left(
1+\frac{\epsilon^2}{2}
\right)
.
\eea

We list the result of regularized HEE for spherical entangling surface in various dimensions
as a series expansion in $\epsilon$  in Table. \ref{table1}. The result for pure AdS corresponds to $M=0$ cases
in Table \ref{table1}, and $M\neq 0$ cases will be discussed in Sec.\ref{3a}.
As it is well known the leading-order divergent terms exhibit the area law. 
When the boundary theory is even-dimensional, {\it i.e.} $d$ is even it also contains logarithmic divergence, whose coefficient is universal and
related to the central charge \cite{Takayanagi:2012kg,Nishioka:2009un,Ryu:2006bv}. 
\begin{table}
\begin{tabular}{c|c}
             & $\text{Area}/(\ell^{d-1}\text{vol}(S^{d-2}))$  \\
\hline\hline
$AdS_3$ & $2\log\left(\frac{1}{\epsilon}\right)+2\log(2\sin\theta_0)+\cdots. $  \\
\hline
$AdS_4$   & $\sin\theta_0\left(\frac{1}{\epsilon}\right)-1+M\frac{\csc\theta_0(1-3\cos\theta_0+\cos 2\theta_0+\sec\theta_0)}{3}+\cdots.$    \\
\hline
$AdS_5$ &  $\frac{\sin^2\theta_0}{2}\left(\frac{1}{\epsilon}\right)^2-\frac{1}{2}\log\left(\frac{1}{\epsilon}\right)
-\frac{\cos 2\theta_0}{12}-\frac{1+3\log2}{6}-\frac{\log(\sin\theta_0)}{2}$
\\
&
$+M\frac{\csc^2\theta_0(3+3\cos\theta_0+6\cos^4\theta_0+3\theta\sec\theta_0\sin 3\theta_0)}{16}
+\cdots.$\\
\hline
$AdS_6$ & $\frac{\sin^3\theta_0}{3}\left(\frac{1}{\epsilon}\right)^3-\frac{\sin\theta_0(5+\cos^2\theta_0)}{6}\left(\frac{1}{\epsilon}\right)+\frac{2}{3}$ 
\\
&
$+M\frac{8}{15}(3+6\cos\theta_0+\cos 2\theta_0)\sec\theta_0\sin^4\frac{\theta_0}{2}\tan\frac{\theta_0}{2}+\cdots.$
\\
\hline
$AdS_7$ & $\frac{\sin^4\theta_0}{4}\left(\frac{1}{\epsilon}\right)^4-\frac{(8+\cos 2\theta_0)\sin^2\theta_0}{12}\left(\frac{1}{\epsilon}\right)^2+\frac{3}{8}\log\left(\frac{1}{\epsilon}\right)
+\frac{3}{16}\log(2\sin^2\theta_0)
+\frac{3}{16}\log 2$  \\
&
$+\frac{11}{1440}\cos 4\theta_0+\frac{17}{180}\cos 2\theta_0+\frac{43}{240}$\\
&
$+M\frac{5+\cos 2\theta_0-6\theta_0\cot\theta_0}{8}+\cdots.$
\\
\hline
\end{tabular}
\caption{HEE for pure $AdS$ and AdS-Schwarzschild black holes in various dimensions}
\label{table1}
\end{table}
\subsection{Metric perturbations and HEE}
We consider small metric perturbations around AdS, and see the change in the calculated
results of HEE. For simplicity and concreteness, we restrict our attention to 
time-dependent but spherically symmetric backgrounds. The form of the metric 
we employ is as follows, 
\bea
ds^2=\frac{\ell^2}{\cos^2x}\biggl(-A(t, x)e^{-2\delta(t, x)}dt^2+A^{-1}(t, x)dx^2+\sin^2x(d\theta^2+\sin^2\theta d\Omega^2_{d-2})\label{metric3}
\biggr) \, . 
\eea
The pure AdS metirc is recovered for $A=1$ and $\delta=0$. 
We note that this metric ansatz was used in e.g. \cite{Bizon:2011gg} for the study of dynamical instability of gravity 
in AdS.  Indeed, 
after we obtain the formulae for perturbed HEE we will employ them to the perturbatively
obtained time-periodic solutions reported in \cite{Maliborski:2013jca,Kim:2014ida,Fodor:2015eia}.

In terms of new variables $ \rho=\sin x, z=\cos \theta$ and treating both $\rho,t$ as functions of $z$, 
the area functional \eqref{area2} is now generalized to 
\bea
\frac{\text{Area}}{\text{vol}(S^{d-2})}=\int dz\frac{\rho^{d-2}(1-z^2)^{\frac{d-3}{2}}}{(1-\rho^2)^{\frac{d-1}{2}}}
\sqrt{-g_{tt}(1-z^2)(t')^2+\frac{g_{xx}(1-z^2)}{1-\rho^2}(\rho')^2+\rho^2},
\label{the area2}
\eea
Here as shorthand we introduced 
\bea
g_{tt}&:=&Ae^{-2\delta}=1-2\sum_{n=1}^\infty\Phi_{n}(t, \rho)\lambda^{n},
\\
g_{xx}&:=&A^{-1}=1+2\sum_{n=1}^\infty\Psi_{n}(t, \rho)\lambda^{n}.
\eea
where $\lambda$ is the perturbation parameter of the metric.
The minimal-area surface for Ryu-Takayanagi formulae is also to be determined perturbatively. 
In general the solutions are expanded in $\lambda$,
\bea
\rho(z)&=&a/z+\sum_{n=1}^\infty\rho_{n}(z)\lambda^{n}, 
\\
t(z)&=&t_0+\sum_{n=1}^\infty t_{n}(z) \lambda^{n}.
\eea
At fixed order of $\lambda$, the configuration functions $\rho_n(z),t_n(z)$ should satisfy certain second-order
inhomogeneous differential equation. One can easily convince oneself that the homogeneous part is independent
of $n$, so the homogeneous solutions $t_{h1},t_{h2}$ and $\rho_{h1},\rho_{h2}$ are the same for all $n$.
Then it is a basic property of ordinary differential equations that particular solutions can be constructed in 
terms of the homogeneous solutions and their Wronskians, $W_t(z)$ and $W_\rho(z)$,  as follows.
\bea
t_{\text{part}}^{(n)}(z)&=&\int_{z_0}^zdz'\frac{(t_{h1}(z')t_{h2}(z)-t_{h1}(z)t_{h2}(z'))G_t^{(n)}(z')}{W_t(z')},
\\
\rho_{\text{part}}^{(n)}(z)&=&\int_{z_0}^zdz'\frac{(\rho_{h1}(z')\rho_{h2}(z)-\rho_{h1}(z)\rho_{h2}(z'))G_\rho^{(n)}(z')}{W_\rho(z')},
\eea
Here $G_t^{(n)}(z)$ and $G_\rho^{(n)}(z)$ are inhomogeneous parts for $t_n(z)$ and $\rho_n(z)$.

To completely determine the solutions $t_n(z),\rho_n(z)$, we need to specify the boundary conditions. From
obvious physical reasoning, we demand they vanish at $z=a$, and have finite values at $z=0$. Another comment
is that since $t(z)$ is trivial at ${\cal O}(\lambda^0)$, the perturbation of $g_{tt}$, or $\Phi_n$, do not incur any
change at the leading nontrivial order ${\cal O}(\lambda)$. This will be shown more explicitly in the following subsections.

\section{Examples}
\subsection{HEE in AdS-Schwarzschild}
Let us now take Schwarzschild black holes in AdS backgrounds. 
The metric is given by
\bea
ds^2=-f_d(r)dt^2+\frac{dr^2}{f_d(r)}+r^2(d\theta^2+\sin^2\theta d\Omega^2_{d-2}), \label{sch_metric}
\eea
where $f_d(r)=1-\frac{M}{r^{d-2}}+\frac{r^2}{\ell^2}$. $M$ is the mass parameter of the black hole.
Following the standard procedure, the black hole mass $M$ is related to the temperature, through
the Wick rotation $\tau=i t$ and requiring the resulting Euclidean geometry to be free from 
a conical singularity. Denoting the position of the horizon by the largest root of $f_d(r_+)=0$,
 the temperature and the entropy of the black hole are given as follows 
\bea
T_{BH}&=&
\frac{\ell^2(d-2)+dr_+^2}{4\pi\ell^2r_+},
\\
S_{BH}&=&\frac{A}{4G_{d+1}}
=\frac{\text{vol}(S^{d-1})}{4G_{d+1}}r_+^{d-1}.
\eea
Since we plan to use perturbation expansion, we assume $M$ is small and
the horizon is  $r_+ \propto M^{1/(d-2)}$.

For $M\neq 0$ the area integral is now 
\bea
\text{Area}=\text{vol}(S^{d-2})\int^1_{\cos^{-1}\theta_0} dz\, \frac{\rho^{d-2}(1-z^2)^{\frac{d-3}{2}}}{(1-\rho^2)^{\frac{d-1}{2}}}
\sqrt{\frac{(1-z^2)(\rho')^2}{(1-\rho^2)(1-M(1-\rho^2)^{d/2}\rho^{2-d})}+\rho^2}.
\label{generalA}
\eea 
Note that since the metric is static the equation for $t(z)$ is trivial and we can set it to a constant.
We are interested in the correction terms to $\text{Area}(\theta_0)$ given for pure AdS in the Table \ref{table1}. 
On general grounds we of course expect that the minimal area surface cannot penetrate 
to the interior of the black hole and there should be a phase transition \cite{Hubeny:2013gta}, 
but here we will compute the small correction terms when the cap-like entanglement subspace is small.
\subsubsection*{Perturbative calculation of HEE for $AdS_4$-Schwarzschild}
\label{3a}
Let us start with the case of $AdS_4$. The integrand of the action functional \eqref{generalA} is now
\bea
L= \frac{\rho}{1-\rho^2}\sqrt{\rho^2+\frac{(1-z^2)}{(1-\rho^2)(1-\frac{M(1-\rho^2)^{3/2}}{\rho})}(\rho')^2} \, . 
\label{s4}
\eea
We treat $M$ as a small parameter and consider a small perturbation away from constant-latitude solution:
\bea
\rho(z)=\frac{a}{z}+M\rho_1(z) + 
\cdots \, . 
\eea

Upon expansion of the equation of motion from \eqref{s4}, one notes that $\rho_1(z)$ should satisfy the following inhomogeneous linear second-order differential equation, 
\bea
\rho_1''&+&\frac{a^2(4-6z^2)+z^2(5z^2-3)}{z(z^2-1)(z^2-a^2)}\rho_1'+\frac{a^2(2-4z^2)+z^2(3z^2-1)}{z^2(z^2-1)(z^2-a^2)}\rho_1
\nn\\ 
&=&\frac{\sqrt{z^2-a^2}(z^2+3z^4-2a^2(1+z^2))}{2z^5(z^2-1)}.
\label{pe4}
\eea
We recall that the general solution of an equation $\rho_1''+P(z)\rho_1'+Q(z)\rho_1=G(z)$ is, in terms of 
homogeneous solutions $u_1,u_2$ and their Wronskian $W=u_1u_2'-u_1'u_2$, given as 
\be
\rho_1(z) = c_1 u_1(z) + c_2 u_2(z) + \int^z_{z_0} dz' \frac{(u_1(z')u_2(z)-u_1(z)u_2(z'))G(z')}{W(z')} . 
\ee
Using the fact that the homogeneous solutions of \eqref{pe4} are 
\bea
u_1&=&1/z ,
\\
u_2&=&\frac{\sqrt{z^2-a^2}}{z^2}+\frac{\sqrt{1-a^2}}{2z}
\nn\\
&+&\frac{1}{2z^2}\log\biggl[
\frac{a^2(2a^2-z^2+z-3)+z^2\sqrt{(1-a^2)(z^2-a^2)}}{a^2(1-z^2)}
\biggr].
\eea
and implementing the boundary condition $\rho_1(z=a)=0$ (in order not to change the position of 
entangling surface at $\theta=\theta_0$), we obtain 
\bea
\rho_1(z)&=&\frac{(a^2-2z+z^2)\sqrt{z^2-a^2}}{2z^3}
-\frac{\sqrt{1-a^2}}{z}\log\biggl[
\frac{a^2+z-\sqrt{(1-a^2)(z^2-a^2)}}{a(1+z)}
\biggr]. \label{solution4}
\eea
Now we evaluate the on-shell value of the action. We introduce a cutoff at  $x=\pi/2-\epsilon$ or equivalently
$z_{min}=a(1+\frac{\epsilon^2}{2})$. The result is that the perturbative terms do not change the divergent part
and gives only the following finite contributions.

One can obviously repeat the above computation in higher dimensional black holes. The results 
are summarized in Table \ref{table1}. 
Note that for small entangling region $a\ra 1 (\theta_0\ra \pi/2)$ and the parts dependent on $M$ vanish. It is natural since in that
case the Ryu-Takayanagi surface is also very small and the correction from the change of the metric should be
negligible. 

\subsection{HEE in quasi-periodic backgrounds}
Let us now turn to time-dependent backgrounds. We will consider here
the spherically-symmetric, time-periodic solutions in the AdS-scalar system, constructed
first in \cite{Maliborski:2013jca} and developed further in \cite{Kim:2014ida,Fodor:2015eia}.
In the metric \eqref{metric3}, a massive scalar field  equation reduces to
\be
\partial_t (e^{\delta} A^{-1} \partial_t \phi ) - 
\frac{1}{\tan^{d-1}x} \partial_x ( A e^{-\delta} \tan^{d-1} x \partial_x \phi )
+\frac{\Delta(\Delta-d)}{\cos^2 x} e^{-\delta} \phi = 0 \, . 
\label{kge}
\ee
Here the mass of the scalar field is $m^2=\Delta(\Delta-d)/\ell^2$. 
The Einstein equation in the presence of matter field excitation reduces to the following equations.
\bea
\delta' &=& -  \sin x \cos x ( A^{-2} e^{2\delta} \dot\phi^2 + 
\phi'^2 ) \, , 
\label{ee1}\\
A' &=& A\delta' + \frac{d-2+2\sin^2x}{\sin x\cos x} (1-A) - 
\frac{\Delta(\Delta-d)\sin x}{\cos x} \phi^2 \, . 
\label{ee2}
\eea

For a perturbative approach, one can first consider a small fluctuation
of the scalar field $\phi\sim {\cal O}(\varepsilon)$ and solve \eqref{kge} in pure AdS, i.e. $A=1,\delta=0$. 
The eigenmodes are in general given in terms of Jacobi polynomials of $u=\cos x$.
In the next step
the scalar configuration is substituted to \eqref{ee1} and \eqref{ee2}. Integrating these equations, one immediately
obtains $A,\delta$ up to ${\cal O}(\varepsilon^2)$. Then this updated background is in turn used in \eqref{kge} to 
get the scalar field up to ${\cal O}(\varepsilon^3)$.  This procedure can be in principle repeated to arbitrarily 
higher orders to construct solutions analytically. One notable feature of this system is resonance and cancellation 
of secular terms. The solutions of \eqref{kge} in pure AdS background are equipped with integer-valued frequency,
and at higher-orders in perturbation the inhomogeneous terms in general can contain secular terms which lead
to linear growth of amplitude in time. However, if one starts with a single mode at ${\cal O}(\varepsilon)$, it turns
out the secular terms always cancel and the only non-trivial effect is the shift of the frequency as a function of 
$\varepsilon$. The requirement that the frequency as a series expansion in $\varepsilon$ should converge puts
an upper limit on the numerical value of $\varepsilon$. It is also worth noting 
that, at least at the first non-trivial order of
nonlinearity ${\cal O}(\varepsilon^3)$, one can show generically certain kinds of secular terms always cancel with each 
other which implies the existence of additional conserved quantities \cite{Craps:2014vaa,Craps:2014jwa}.

We can use the area integral in \eqref{the area2}, with the identification of perturbation parameter $\lambda=\varepsilon^2$.
Among the solutions reported in \cite{Kim:2014ida}, we will for definiteness consider $AdS_5$ case. When the scalar field is massless for instance, at ${\cal O}(\lambda)$ the metric is given as ($u=\cos x=\sqrt{1-\rho^2}$)
\bea
\delta(t, u)&=&\lambda \left[-1+u^8+\frac{3\cos(8t)}{5}+u^8\cos(8t)-\frac{8u^{10}\cos(8t)}{5} \right],
\\
A(t, u)&=&1-\lambda\left[\frac{2u^4}{3}+\frac{2u^6}{3}-\frac{4u^8}{3}-2u^8\cos(8t)+2u^{10}\cos(8t)\right] .
\eea
Then the equation for $\rho_1(z)$ is
\bea
\rho''_1(z)&+&\frac{4a^2-7a^2z^2-2z^2+5z^4}{z(1-z^2)(a^2-z^2)}\rho'_1(z)+\frac{3z^4+2a^2-5a^2z^2}{z^2(1-z^2)(a^2-z^2)}\rho_1(z)
\nn\\
&=&\frac{2a^3(a^2-z^2)^2(-8a^2z^2+9z^4+2a^2z^4-3(3z^4+2a^2z^2(z^2-4)-a^4(2z^2-5))\cos(8t_0))}{3z^{13}(-1+z^2)}.
\nn\\
\eea
When we solve this equation with appropriate boundary condition, we find
\bea
\rho_1(z)&=&\frac{a(a^2-z^2)^2(36z^4-a^2z^2(33+25z^2)+2a^4(5+3z^2+3z^4))}{105z^9}
\nn\\
&+&\frac{a(a^2-z^2)^2}{315z^{11}}\biggl(-82z^6+a^2z^4(151+95z^2)-2a^4z^2(60+31z^2+32z^4)
\nn\\
&+&a^6(35+15z^2+16z^4+16z^6)
\biggr)\cos(8t_0).
\eea
Note that the minimal area configuration now contains dependence on boundary time $t_0$. 

We can do the same computation with $t_1(z)$. 
\bea
t''_1(z)+\frac{3z^4+2a^2-5a^2z^2}{z(1-z^2)(a^2-z^2)}t'_1(z)=-\frac{8a^4(z^2-a^2)^3}{z^{12}}\sin(8t_0).
\eea
One notable difference from the equation for $\rho_1(z)$ is that there is no linear term in $t_1(z)$. This
means one of the homogeneous solutions is just constant. The solution with the correct boundary condition 
turns out to be
\bea
t_1(z)&=&-\frac{(a^2-z^2)^2}{315z^{10}}\biggl(-19z^6+19a^2z^4(-2+5z^2)+a^4(69z^2-62z^4-64z^6)
\nn\\
&+&a^6(-28+15z^2+16z^4+16z^6)\biggr)\sin(8t_0).
\eea
Now we can substitute these results into the expression for HEE. The result is
\begin{itemize}
\item Massless scalar field ($\Delta=4$)
\bea
\frac{\text{Area}}{\ell^3\text{vol}(S^2)}&=&\frac{1-a^2}{2}\biggl(\frac{1}{\epsilon}\biggr)^2-\frac{1}{2}\log\biggl(\frac{1}{\epsilon}\biggr)-\frac{1+2\log 2+\log(1-a^2)}{4}
\nn\\
&+&\lambda\biggl[\frac{2(1-a^2)^2}{105}\biggl(12-5a^2-4(1-a^2)^2\cos(8t_0)\biggr)\biggr]+\cdots,
\label{as1}
\eea
\end{itemize}
where we again introduced a cutoff $\epsilon$ at the boundary $z=a$ to regularize the area.
We have also calculated the entanglement entropy for massive scalar fields. Here we will just record the perturbation results at order ${\cal O}(\lambda)$.
\begin{itemize}
\item $\Delta=3$
\bea
\delta_\lambda\biggl(\frac{\text{Area}}{\ell^3\text{vol}(S^2)}\biggr)&=&
\frac{3(1-a^2)^2}{140}\biggl(7-5(1-a^2)\cos(6t_0)\biggr)+\cdots. \label{as2}
\eea
\item $\Delta=2$
\bea
\delta_\lambda\biggl(\frac{\text{Area}}{\ell^3\text{vol}(S^2)}\biggr)&=&\frac{(1-a^2)^2}{5}\cos(4t_0)+\cdots. \label{as3}
\eea
\end{itemize}

\begin{figure}[ht]
\centering
    \begin{subfigure}[b]{0.44\textwidth}
        \includegraphics[width=80mm]{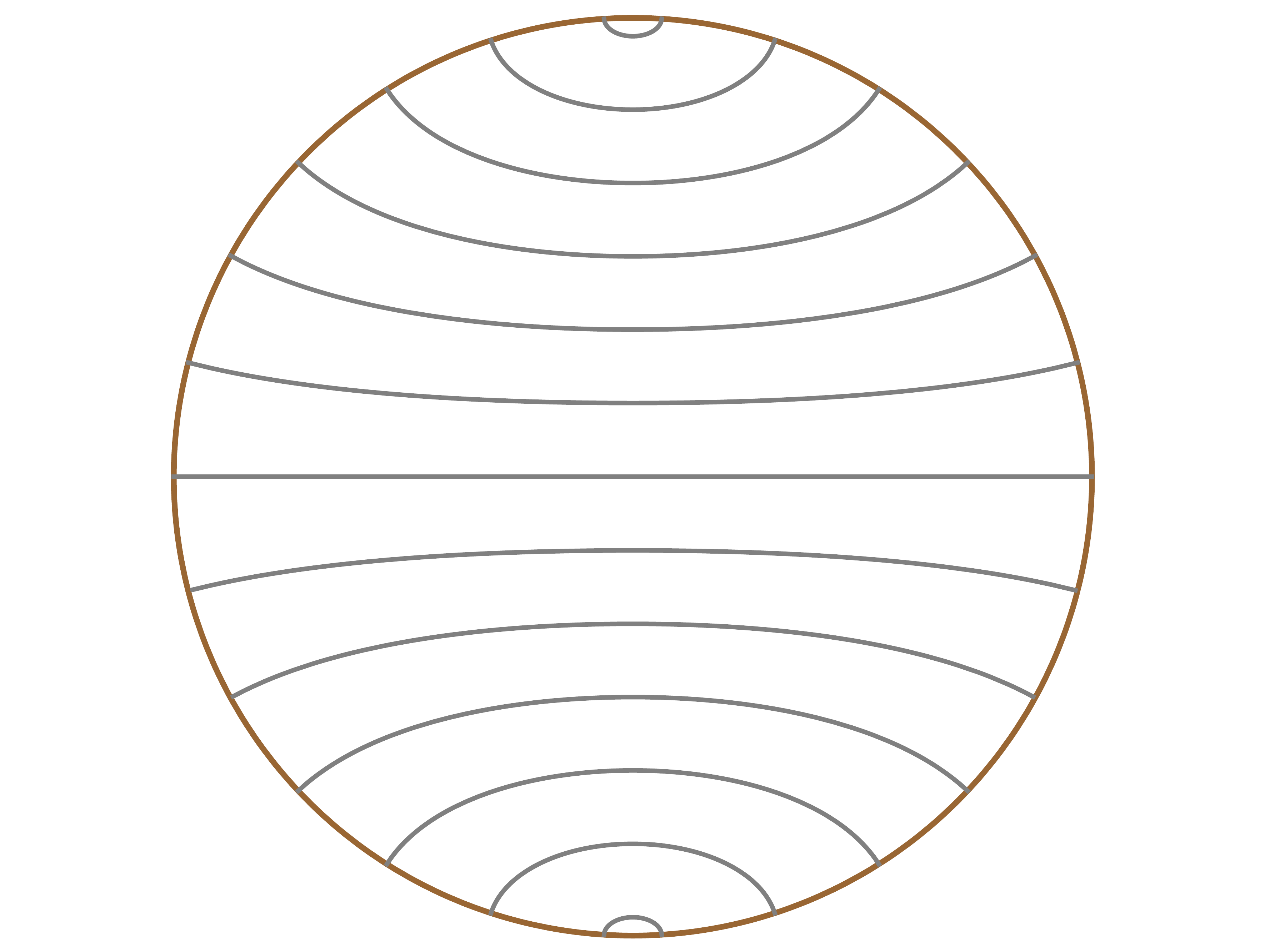}
       \caption{AdS}
        \label{constlat}
    \end{subfigure}
    \qquad
\centering
    \begin{subfigure}[b]{0.44\textwidth}
        \includegraphics[width=80mm]{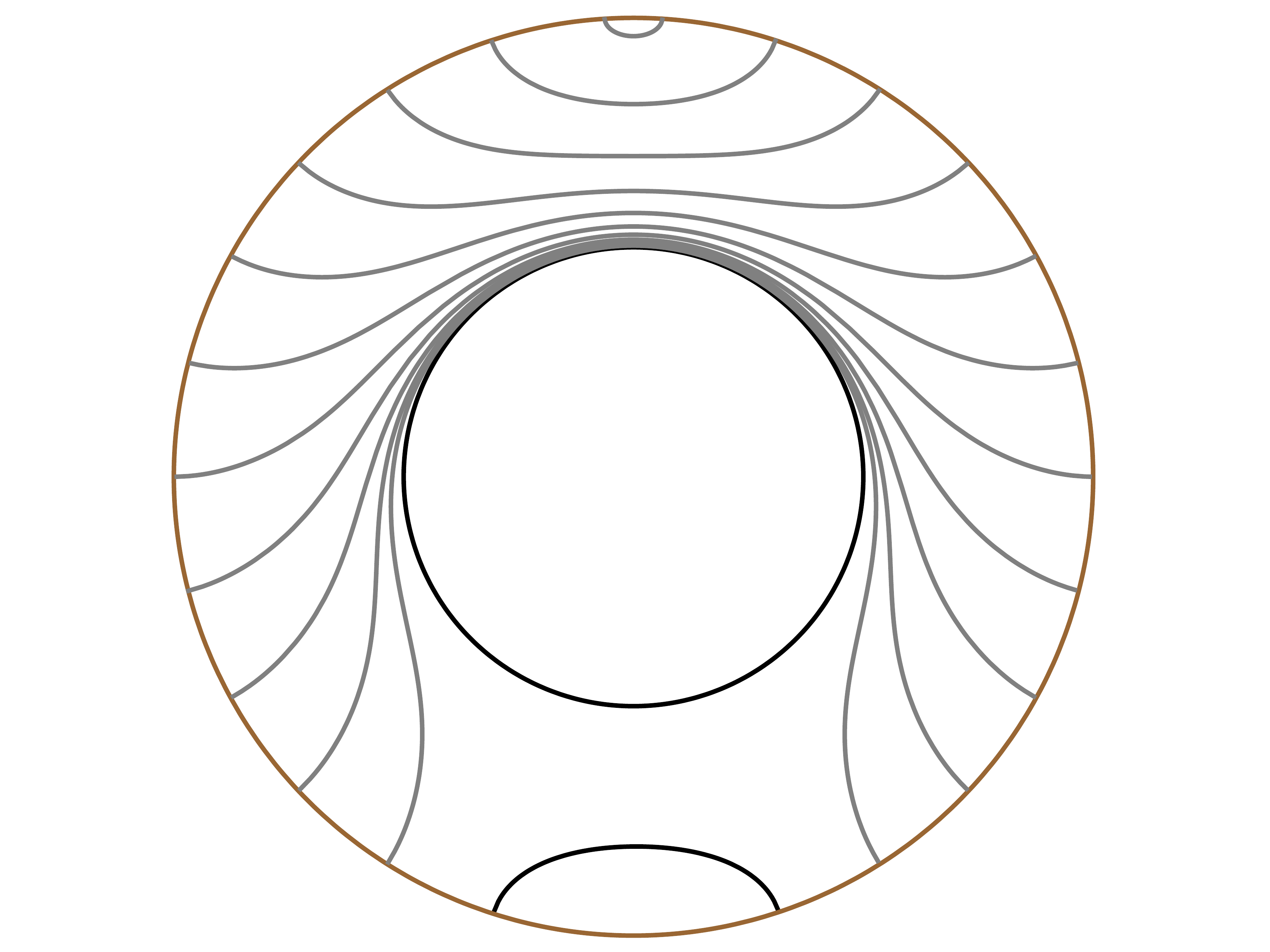}
        \caption{AdS-Schwarzschild}
    \end{subfigure}
    \caption{Minimal-area surfaces.}
\label{}
\end{figure}
\begin{figure}[ht]
\centering
    \begin{subfigure}[b]{0.44\textwidth}
        \includegraphics[width=80mm]{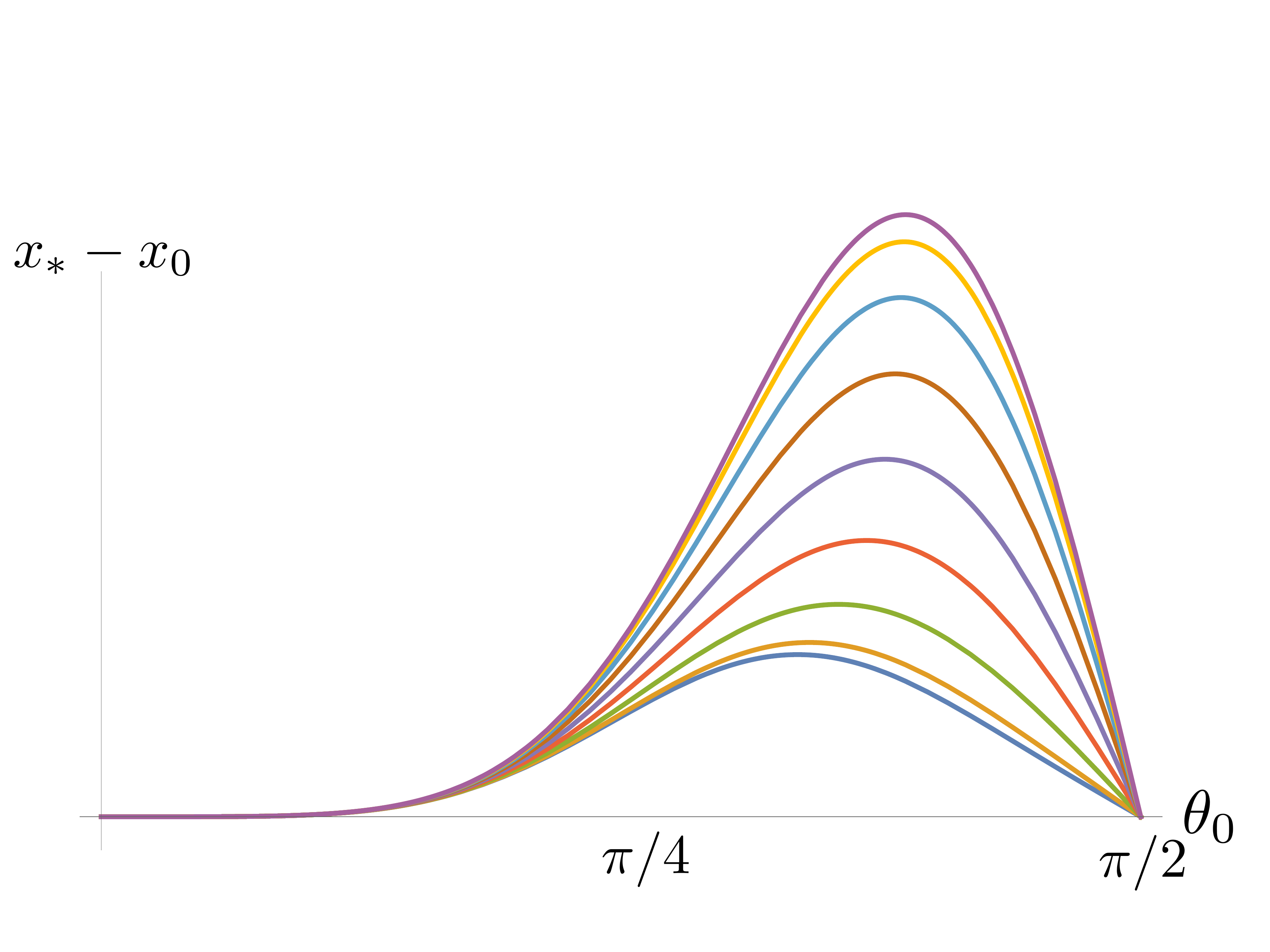}
       \caption{$\Delta=4$}
    \end{subfigure}
    \qquad
\centering
    \begin{subfigure}[b]{0.44\textwidth}
        \includegraphics[width=80mm]{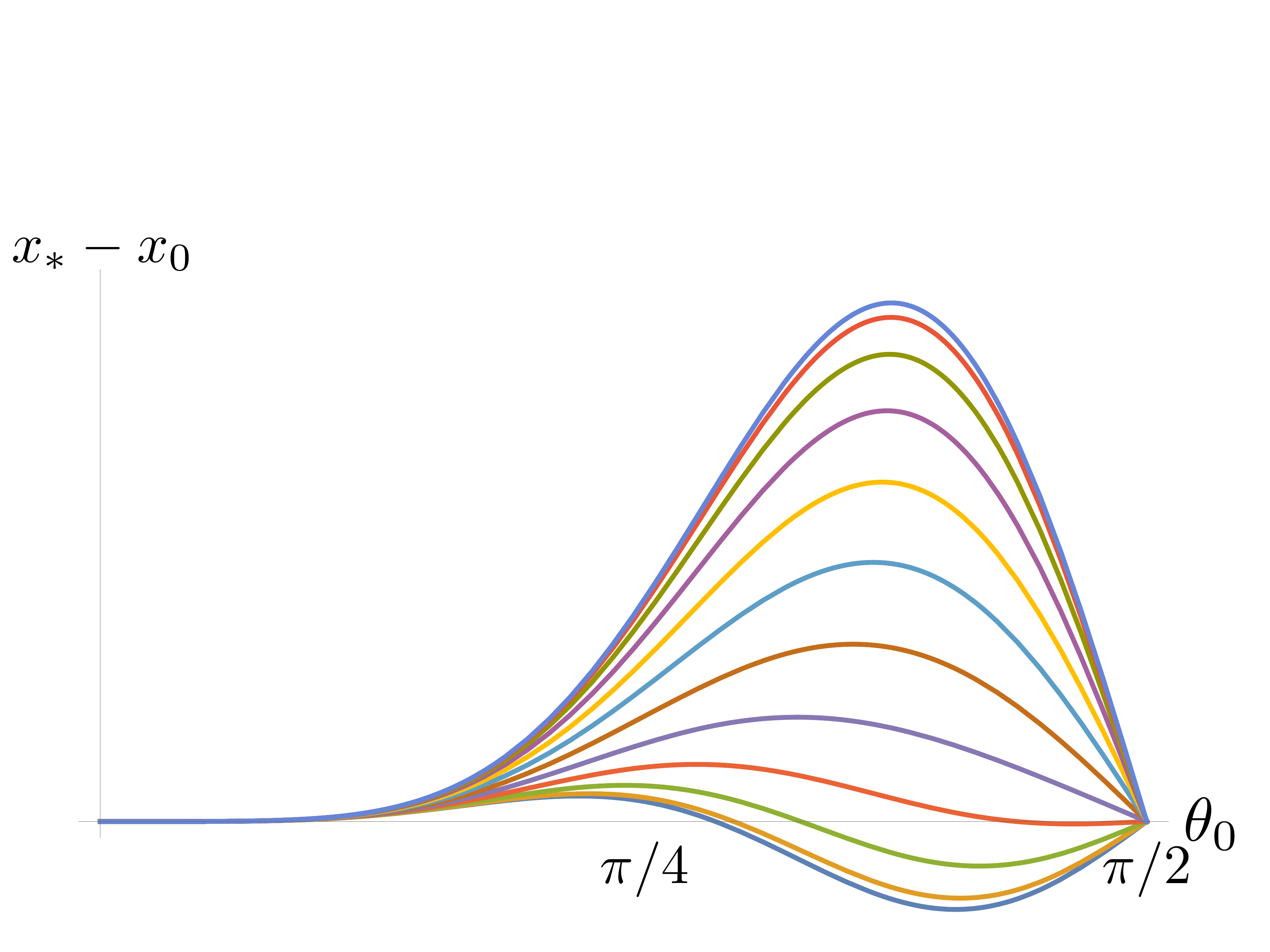}
        \caption{$\Delta=3$}
    \end{subfigure}
    \caption{The change of the point at $\theta=0$ at different times.}
\label{rttime}
\end{figure}
To help understanding the change of  the minimal-area surface in time, we plotted the change of the point at $\theta=0$
as a function of time in Fig.\ref{rttime}. It is obvious from Fig.\ref{constlat} that for each curve, $\theta=0$ is the closest point to the center of AdS. Holographically speaking, $\theta=0$ points are thus the most IR point we can probe using HEE. Note that
the perturbative computations are valid for small $\theta_0$ region, i.e. when the cap-like regions are small. 
The curves in Fig.\ref{rttime} represent how much the HEE curves are pushed away from the AdS origin, compared to 
the constant-latitude curve \eqref{cl}, due to 
the stress-energy tensor turned on by the nontrivial scalar field configuration. Another way of 
interpreting the time-dependence of our metric is that it should be at least qualitatively similar to the formation 
of black hole, due to a collapse of spherical shell. What we see in Fig.\ref{rttime} is the change of HEE in time, 
as a spherically symmetric shell starts from AdS infinity and shrinking to a point at the AdS center. We clearly see
in Fig.\ref{rttime}
that $x_*-x_0$ is smaller for $\Delta=3$, when compared to $\Delta=4$ case.
It agrees with our intuition that when the scalar field is less massive (and more tachyonic), the effect on 
the gravitational system should be weaker and eventually the push-away effect should be also weaker.
\section{Entanglement Temperature}
One encounters a relation similar to the thermodynamical first law from the entanglement entropy in a small
subsystem with excitations  \cite{Bhattacharya:2012mi,Allahbakhshi:2013rda,Chakraborty:2014lfa,Dey:2014voa,Chakraborty:2014lfa,Momeni:2015vka,Ghoroku:2015apa,Mishra:2015cpa,Pal:2015mda,Matsueda:2015bba,Li:2015ola,Mansoori:2015sit,Park:2015hcz}. Namely, the change of entanglement entropy and the energy are linearly related, and the entanglement temperature is defined as
$T_{\text{ent}}\Delta S_A=\Delta E_A$. Since we
consider small fluctuations and small entangling regions in this paper, let us also compute the entanglement temperature. 
We already computed $\Delta S_A$, so we only need to compute the change of energy $\Delta E_A$. 
 It turns out that, for a small entangling region $A$, the entanglement temperature $T_{\text{ent}}$ is proportional to the inverse size of the subsystem,
and the proportionality coefficient is determined by the shape of the entangling region and independent of the strength of
the fluctuation. We take $AdS_5$ case for concreteness, but obviously the computation can be easily repeated in other
dimensions.

\subsection{AdS-Schwarzschild black hole}
In terms of Fefferman-Graham coordinates, asymptotically AdS metric is written as $ds^2=\frac{dz^2}{z^2}+\gamma_{ab}dx^adx^b$. Then \eqref{sch_metric} is rewritten as follows \cite{Tetradis:2009bk},
\bea
ds^2=\frac{1}{z^2}\biggl[dz^2-\frac{(1-\alpha_1z^4)^2}{1+\alpha_2 z^2+\alpha_3 z^4}dt^2+(1+\alpha_2 z^2+\alpha_3 z^4)(d\theta^2+\sin^2\theta\Omega_2^2)\biggr],
\label{fg}
\eea
where $\alpha_1=1, \alpha_2=-\frac{1}{2}$, and $\alpha_3=\frac{1+4 M}{16}$. The new holographic
coordinate and $z$ is defined by
\bea
\frac{dz}{z}=-\frac{dr}{\sqrt{f(r)}}
\Longrightarrow
z^2=\frac{4}{1+2r^2+2r\sqrt{f(r)}}.
\eea
Expanding the induced metric $\gamma_{ab}$ near the AdS boundary ($z=0$), one gets
\bea
\gamma_{ab}=\gamma_{(0)ab}+z^2\gamma_{(2)ab}+z^4\gamma_{(4)ab}+\cdots.
\eea
Then the stress tensor of the boundary theory is given by \cite{deHaro:2000xn,Skenderis:2000in}
\bea
T_{ab}^{\text{CFT}}=\frac{4}{16\pi G_N}\biggl[
\gamma_{(4)ab}-\frac{1}{8}\gamma_{(0)ab}[(\text{Tr}\gamma_{(2)})^2-\text{Tr}\gamma_{(2)}^2]
-\frac{1}{2}(\gamma_{(2)}^2)_{ab}+\frac{1}{4}\gamma_{(2)ab}\text{Tr}\gamma_{(2)}\biggr],
\eea
which upon substitution of \eqref{fg} becomes
\bea
T_{tt}=\frac{3\ell^3}{64\pi G_N}+M\frac{3\ell^3}{16\pi G_N}.
\eea
Here the first term on the right-hand side is the time-time component of the stress tensor of pure five-dimensional AdS spacetime (Casimir energy), and the second term is fluctuation energy for small black hole mass. At $r=\infty$ $(z=0)$ the metric (\ref{sch_metric}) the boundary metric approaches $R\times S^3$. The energy is given by
\bea
\Delta E_A&=&\text{Vol}(S^2)\int^{\theta_0}_0 d\theta \sin^2\theta \times\Delta T_{tt}
\nonumber\\
&=&\frac{3(\cos^{-1}a-a\sqrt{1-a^2})}{8G_5}M,
\eea
where $a=\cos\theta_0$ and we have set $\ell=1$. Combining $\Delta E_A$ and $\Delta S_A$ and expand for small $\theta_0$ angle, we obtain 
\bea
\frac{\Delta S_A}{\Delta E_A}=\frac{1}{T_{\text{ent}}}=\frac{4\pi}{5}
\theta_0.
\label{entT}
\eea
\subsection{AdS-Scalar systems}
To compute the boundary stress tensor for AdS-scalar systems we employ the method used in \cite{Myers:1999psa}. Let  $\mathcal{M}$ be the spacetime manifold with time-like boundary $\partial\mathcal{M}$. The induced metric on the boundary is $\gamma_{\mu\nu}=g_{\mu\nu}-n_\mu n_\nu$, where $n^\mu$ is the outward-pointing normal vector normal to $\partial\mathcal{M}$. The extrinsic curvature on $\partial\mathcal{M}$ is then $\Theta_{\mu\nu}=-\gamma_{\mu}^{\;\;\rho}\nabla_\rho n_\nu$. 

In five-dimensions, the Einstein action including the boundary term is given by
\bea
I=\frac{1}{16\pi G_5}\int_{\mathcal{M}}d^5x\sqrt{-g}(R-2\Lambda)-\frac{1}{8\pi G_5}\oint_{\partial\mathcal{M}}d^4x\sqrt{-\gamma}\Theta+I_{\text{matter}},\label{cov_action}
\eea
where $\Theta$ is the trace of the extrinsic curvature of the boundary. 
The boundary stress tensor is given by
\bea
\tau^{ab}\equiv\frac{2}{\sqrt{-\gamma}}\frac{\delta I}{\delta\gamma_{ab}}=\frac{1}{8\pi G_N}(\Theta^{ab}-\gamma^{ab}\Theta^c_c).\label{bd_tensor}
\eea
We need to regularize this action, and introduce a reference background metric $g^{0}_{\mu\nu}$ which asymptotically agrees with $g_{\mu\nu}$ on the boundary. The regularized action in this subtraction scheme is given by  $\hat{I}=I(g)-I^0(g^0)$ and a finite stress tensor is given by \cite{Hawking:1995fd}
\bea
\hat{\tau}^{ab}\equiv\frac{2}{\sqrt{-\gamma}}\frac{\delta \hat{I}}{\delta\gamma_{ab}}=\tau^{ab}-(\tau^0)^{ab} . 
\label{reg_em}
\eea
The stress tensor $\langle T \rangle_{ab}$ in the field theory is given, following the standard AdS/CFT dictionary, as 
\bea
\sqrt{-h}h^{ab}\langle T\rangle_{bc}=\lim_{R\rightarrow\infty}\sqrt{-\gamma}\gamma^{ab}\hat{\tau}_{bc},\label{field_tensor}
\eea
where $h_{ab}$ is the background metric for the field theory spacetime. 

For our ansatz \eqref{metric3},  
the normal vector to the surface at $x=x_0$ is
\bea
n^\mu=\frac{\cos x_0}{\ell \sqrt{1+\lambda\Psi(t,x_0)}}\delta^\mu_x . 
\eea
In the standard ADM decomposion of the metric we read off 
\bea
\gamma_{00}=-\frac{\ell^2(1+\lambda\Phi(t,x_0))}{\cos^2x_0},~~~~\gamma_{ij}=\ell^2\tan^2 x_0\bar{\gamma}_{ij},
\eea
where $i,j=1,2,3$ and  $\bar{\gamma}_{ij}$ is the metric on a unit 3-sphere. The computation is then
straightforward, and we obtain 
\bea
\hat{\tau}_{tt}=\frac{\ell\cos^2 x_0\lambda}{8\pi G_5}+\cdots . 
\eea
Finally using the relation \eqref{field_tensor} one arrives at 
\bea
\langle T_{00}\rangle=\frac{\ell^3\lambda}{8\pi G_5} , 
\eea 
which obviously gives the same entanglement temperature as \eqref{entT}.
For the massive scalar fields $\Delta=3, 2$ we have checked that the results are always the same as \eqref{entT},
verifying the statement in  \cite{Bhattacharya:2012mi} that the proportionality coefficient for entanglement entropy as a function of the subsystem size should be universal and does not depend on the details of the excitations.

\section{Discussions}
In this paper, we have calculated the holographic entanglement entropy (HEE) perturbatively in the
backgrounds which are small deformations of AdS vacuum. More  concretely, we considered Schwarzschild black holes, and also harmonically driven time-dependent backgrounds. We chose the entanglement region to be cap-like ones around north pole of the boundary space.
The motivation of this work was to see how small excitations in the gravity side manifest itself in HEE. Our observation is that
the metric perturbation around the AdS vacuum does not affect the divergent terms and the change is in the finite part, as one
expands with respect to the regularization parameter.
We also considered the entanglement thermodynamics and computed
the entanglement temperature for small sub-systems.
 This result is
consistent with Ref.\cite{Bhattacharya:2012mi},
where it was argued that  the entanglement temperature should exhibit a universal feature which is proportional to the inverse size of the system. We have used the methods proposed in \cite{Myers:1999psa,deHaro:2000xn}
to compute the boundary stress tensor and obtained the thermodynamic first-law like relation. We have checked that the proportional factor has always the same value for the system considered in this paper.

As a final comment, it would be also interesting to extend the study of time-dependent background and compute
another thermodynamic quantity, the so-called entanglement pressure, which further generalizes the first law of thermodynamics for quantum entanglement \cite{Allahbakhshi:2013rda}.

\begin{acknowledgments}
This work was supported by National Research Foundation of Korea 
(NRF) grants funded by the Korea government (MEST) with grant No. 2015R1D1A1A09059301.
\end{acknowledgments}
\bibliography{ht}

\providecommand{\href}[2]{#2}\begingroup\raggedright\begin{thebibliography}{10}

\bibitem{Hooft:1993gx}
G.~'t~Hooft, {\it {Dimensional reduction in quantum gravity}},  in {\em
  {Salamfest 1993:0284-296}}, pp.~0284--296, 1993.
\newblock \href{http://arxiv.org/abs/gr-qc/9310026}{{\tt gr-qc/9310026}}.

\bibitem{Susskind:1994vu}
L.~Susskind, {\it {The World as a hologram}},  {\em J. Math. Phys.} {\bf 36}
  (1995) 6377--6396, [\href{http://arxiv.org/abs/hep-th/9409089}{{\tt
  hep-th/9409089}}].

\bibitem{Maldacena:1997re}
J.~M. Maldacena, {\it {The Large N limit of superconformal field theories and
  supergravity}},  {\em Adv.Theor.Math.Phys.} {\bf 2} (1998) 231--252,
  [\href{http://arxiv.org/abs/hep-th/9711200}{{\tt hep-th/9711200}}].

\bibitem{Ryu:2006bv}
S.~Ryu and T.~Takayanagi, {\it {Holographic derivation of entanglement entropy
  from AdS/CFT}},  {\em Phys. Rev. Lett.} {\bf 96} (2006) 181602,
  [\href{http://arxiv.org/abs/hep-th/0603001}{{\tt hep-th/0603001}}].

\bibitem{Swingle:2009bg}
B.~Swingle, {\it {Entanglement Renormalization and Holography}},  {\em Phys.
  Rev.} {\bf D86} (2012) 065007, [\href{http://arxiv.org/abs/0905.1317}{{\tt
  arXiv:0905.1317}}].

\bibitem{Vidal:2007hda}
G.~Vidal, {\it {Entanglement Renormalization}},  {\em Phys. Rev. Lett.} {\bf
  99} (2007), no.~22 220405, [\href{http://arxiv.org/abs/cond-mat/0512165}{{\tt
  cond-mat/0512165}}].

\bibitem{Hubeny:2007xt}
V.~E. Hubeny, M.~Rangamani, and T.~Takayanagi, {\it {A Covariant holographic
  entanglement entropy proposal}},  {\em JHEP} {\bf 0707} (2007) 062,
  [\href{http://arxiv.org/abs/0705.0016}{{\tt arXiv:0705.0016}}].

\bibitem{Balasubramanian:2010ce}
V.~Balasubramanian, A.~Bernamonti, J.~de~Boer, N.~Copland, B.~Craps,
  E.~Keski-Vakkuri, B.~Muller, A.~Schafer, M.~Shigemori, and W.~Staessens, {\it
  {Thermalization of Strongly Coupled Field Theories}},  {\em Phys. Rev. Lett.}
  {\bf 106} (2011) 191601, [\href{http://arxiv.org/abs/1012.4753}{{\tt
  arXiv:1012.4753}}].

\bibitem{Balasubramanian:2011ur}
V.~Balasubramanian, A.~Bernamonti, J.~de~Boer, N.~Copland, B.~Craps,
  E.~Keski-Vakkuri, B.~Muller, A.~Schafer, M.~Shigemori, and W.~Staessens, {\it
  {Holographic Thermalization}},  {\em Phys. Rev.} {\bf D84} (2011) 026010,
  [\href{http://arxiv.org/abs/1103.2683}{{\tt arXiv:1103.2683}}].

\bibitem{Balasubramanian:2011at}
V.~Balasubramanian, A.~Bernamonti, N.~Copland, B.~Craps, and F.~Galli, {\it
  {Thermalization of mutual and tripartite information in strongly coupled two
  dimensional conformal field theories}},  {\em Phys. Rev.} {\bf D84} (2011)
  105017, [\href{http://arxiv.org/abs/1110.0488}{{\tt arXiv:1110.0488}}].

\bibitem{Allais:2011ys}
A.~Allais and E.~Tonni, {\it {Holographic evolution of the mutual
  information}},  {\em JHEP} {\bf 01} (2012) 102,
  [\href{http://arxiv.org/abs/1110.1607}{{\tt arXiv:1110.1607}}].

\bibitem{Keranen:2011xs}
V.~Keranen, E.~Keski-Vakkuri, and L.~Thorlacius, {\it {Thermalization and
  entanglement following a non-relativistic holographic quench}},  {\em Phys.
  Rev.} {\bf D85} (2012) 026005, [\href{http://arxiv.org/abs/1110.5035}{{\tt
  arXiv:1110.5035}}].

\bibitem{Caceres:2012em}
E.~Caceres and A.~Kundu, {\it {Holographic Thermalization with Chemical
  Potential}},  {\em JHEP} {\bf 09} (2012) 055,
  [\href{http://arxiv.org/abs/1205.2354}{{\tt arXiv:1205.2354}}].

\bibitem{Baron:2012fv}
W.~Baron, D.~Galante, and M.~Schvellinger, {\it {Dynamics of holographic
  thermalization}},  {\em JHEP} {\bf 1303} (2013) 070,
  [\href{http://arxiv.org/abs/1212.5234}{{\tt arXiv:1212.5234}}].

\bibitem{Hubeny:2013hz}
V.~E. Hubeny, M.~Rangamani, and E.~Tonni, {\it {Thermalization of Causal
  Holographic Information}},  {\em JHEP} {\bf 05} (2013) 136,
  [\href{http://arxiv.org/abs/1302.0853}{{\tt arXiv:1302.0853}}].

\bibitem{Liu:2013iza}
H.~Liu and S.~J. Suh, {\it {Entanglement Tsunami: Universal Scaling in
  Holographic Thermalization}},  {\em Phys. Rev. Lett.} {\bf 112} (2014)
  011601, [\href{http://arxiv.org/abs/1305.7244}{{\tt arXiv:1305.7244}}].

\bibitem{Liu:2013qca}
H.~Liu and S.~J. Suh, {\it {Entanglement growth during thermalization in
  holographic systems}},  {\em Phys. Rev.} {\bf D89} (2014), no.~6 066012,
  [\href{http://arxiv.org/abs/1311.1200}{{\tt arXiv:1311.1200}}].

\bibitem{Hubeny:2013dea}
V.~E. Hubeny and H.~Maxfield, {\it {Holographic probes of collapsing black
  holes}},  {\em JHEP} {\bf 03} (2014) 097,
  [\href{http://arxiv.org/abs/1312.6887}{{\tt arXiv:1312.6887}}].

\bibitem{Alishahiha:2014jxa}
M.~Alishahiha, M.~R.~M. Mozaffar, and M.~R. Tanhayi, {\it {On the Time
  Evolution of Holographic n-partite Information}},  {\em JHEP} {\bf 09} (2015)
  165, [\href{http://arxiv.org/abs/1406.7677}{{\tt arXiv:1406.7677}}].

\bibitem{Bai:2014tla}
X.~Bai, B.-H. Lee, L.~Li, J.-R. Sun, and H.-Q. Zhang, {\it {Time Evolution of
  Entanglement Entropy in Quenched Holographic Superconductors}},
  \href{http://arxiv.org/abs/1412.5500}{{\tt arXiv:1412.5500}}.

\bibitem{daSilva:2014zva}
E.~da~Silva, E.~Lopez, J.~Mas, and A.~Serantes, {\it {Collapse and Revival in
  Holographic Quenches}},  {\em JHEP} {\bf 04} (2015) 038,
  [\href{http://arxiv.org/abs/1412.6002}{{\tt arXiv:1412.6002}}].

\bibitem{Ugajin:2014nca}
Y.~Nakaguchi, N.~Ogawa, and T.~Ugajin, {\it {Holographic Entanglement and
  Causal Shadow in Time-Dependent Janus Black Hole}},
  \href{http://arxiv.org/abs/1412.8600}{{\tt arXiv:1412.8600}}.

\bibitem{Keranen:2015fqa}
V.~Keranen, H.~Nishimura, S.~Stricker, O.~Taanila, and A.~Vuorinen, {\it
  {Gravitational collapse of thin shells: Time evolution of the holographic
  entanglement entropy}},  {\em JHEP} {\bf 06} (2015) 126,
  [\href{http://arxiv.org/abs/1502.01277}{{\tt arXiv:1502.01277}}].

\bibitem{Ziogas:2015aja}
V.~Ziogas, {\it {Holographic mutual information in global Vaidya-BTZ
  spacetime}},  {\em JHEP} {\bf 09} (2015) 114,
  [\href{http://arxiv.org/abs/1507.00306}{{\tt arXiv:1507.00306}}].

\bibitem{Leichenauer:2015nxa}
S.~Leichenauer, {\it {Thermal Corrections to Entanglement Entropy from
  Holography}},  \href{http://arxiv.org/abs/1502.07348}{{\tt
  arXiv:1502.07348}}.

\bibitem{Mishra:2015cpa}
R.~Mishra and H.~Singh, {\it {Perturbative entanglement thermodynamics for AdS
  spacetime: Renormalization}},  \href{http://arxiv.org/abs/1507.03836}{{\tt
  arXiv:1507.03836}}.

\bibitem{Maliborski:2013jca}
M.~Maliborski and A.~Rostworowski, {\it {Time-Periodic Solutions in an Einstein
  AdS--Massless-Scalar-Field System}},  {\em Phys. Rev. Lett.} {\bf 111} (2013)
  051102, [\href{http://arxiv.org/abs/1303.3186}{{\tt arXiv:1303.3186}}].

\bibitem{Kim:2014ida}
N.~Kim, {\it {Time-periodic solutions of massive scalar fields in dynamical AdS
  background: Perturbative constructions}},  {\em Phys.Lett.} {\bf B742} (2015)
  274--278, [\href{http://arxiv.org/abs/1411.1633}{{\tt arXiv:1411.1633}}].

\bibitem{Bakas:2015opa}
I.~Bakas and G.~Pastras, {\it {Entanglement Entropy and Duality in AdS(4)}},
  \href{http://arxiv.org/abs/1503.00627}{{\tt arXiv:1503.00627}}.

\bibitem{Takayanagi:2012kg}
T.~Takayanagi, {\it {Entanglement Entropy from a Holographic Viewpoint}},  {\em
  Class. Quant. Grav.} {\bf 29} (2012) 153001,
  [\href{http://arxiv.org/abs/1204.2450}{{\tt arXiv:1204.2450}}].

\bibitem{Nishioka:2009un}
T.~Nishioka, S.~Ryu, and T.~Takayanagi, {\it {Holographic Entanglement Entropy:
  An Overview}},  {\em J. Phys.} {\bf A42} (2009) 504008,
  [\href{http://arxiv.org/abs/0905.0932}{{\tt arXiv:0905.0932}}].

\bibitem{Bizon:2011gg}
P.~Bizon and A.~Rostworowski, {\it {On weakly turbulent instability of anti-de
  Sitter space}},  {\em Phys.Rev.Lett.} {\bf 107} (2011) 031102,
  [\href{http://arxiv.org/abs/1104.3702}{{\tt arXiv:1104.3702}}].

\bibitem{Fodor:2015eia}
G.~Fodor, P.~Forg{\'a}cs, and P.~Grandcl{\'e}ment, {\it {Self-gravitating
  scalar breathers with negative cosmological constant}},
  \href{http://arxiv.org/abs/1503.07746}{{\tt arXiv:1503.07746}}.

\bibitem{Hubeny:2013gta}
V.~E. Hubeny, H.~Maxfield, M.~Rangamani, and E.~Tonni, {\it {Holographic
  entanglement plateaux}},  {\em JHEP} {\bf 1308} (2013) 092,
  [\href{http://arxiv.org/abs/1306.4004}{{\tt arXiv:1306.4004}}].

\bibitem{Craps:2014vaa}
B.~Craps, O.~Evnin, and J.~Vanhoof, {\it {Renormalization group, secular term
  resummation and AdS (in)stability}},  {\em JHEP} {\bf 1410} (2014) 48,
  [\href{http://arxiv.org/abs/1407.6273}{{\tt arXiv:1407.6273}}].

\bibitem{Craps:2014jwa}
B.~Craps, O.~Evnin, and J.~Vanhoof, {\it {Renormalization, averaging,
  conservation laws and AdS (in)stability}},  {\em JHEP} {\bf 1501} (2015) 108,
  [\href{http://arxiv.org/abs/1412.3249}{{\tt arXiv:1412.3249}}].

\bibitem{Bhattacharya:2012mi}
J.~Bhattacharya, M.~Nozaki, T.~Takayanagi, and T.~Ugajin, {\it {Thermodynamical
  Property of Entanglement Entropy for Excited States}},  {\em Phys. Rev.
  Lett.} {\bf 110} (2013), no.~9 091602,
  [\href{http://arxiv.org/abs/1212.1164}{{\tt arXiv:1212.1164}}].

\bibitem{Allahbakhshi:2013rda}
D.~Allahbakhshi, M.~Alishahiha, and A.~Naseh, {\it {Entanglement
  Thermodynamics}},  {\em JHEP} {\bf 1308} (2013) 102,
  [\href{http://arxiv.org/abs/1305.2728}{{\tt arXiv:1305.2728}}].

\bibitem{Chakraborty:2014lfa}
S.~Chakraborty, P.~Dey, S.~Karar, and S.~Roy, {\it {Entanglement thermodynamics
  for an excited state of Lifshitz system}},  {\em JHEP} {\bf 04} (2015) 133,
  [\href{http://arxiv.org/abs/1412.1276}{{\tt arXiv:1412.1276}}].

\bibitem{Dey:2014voa}
A.~Dey, S.~Mahapatra, and T.~Sarkar, {\it {Very General Holographic
  Superconductors and Entanglement Thermodynamics}},  {\em JHEP} {\bf 12}
  (2014) 135, [\href{http://arxiv.org/abs/1409.5309}{{\tt arXiv:1409.5309}}].

\bibitem{Momeni:2015vka}
D.~Momeni, M.~Raza, H.~Gholizade, and R.~Myrzakulov, {\it {Realization of
  Holographic Entaglement Temperature for a Nearly-AdS Boundary}},
  \href{http://arxiv.org/abs/1505.00215}{{\tt arXiv:1505.00215}}.

\bibitem{Ghoroku:2015apa}
K.~Ghoroku and M.~Ishihara, {\it {Entanglement temperature for the excitation
  of SYM theory in the (de)confinement phase}},  {\em Phys. Rev.} {\bf D92}
  (2015), no.~8 085017, [\href{http://arxiv.org/abs/1506.06474}{{\tt
  arXiv:1506.06474}}].

\bibitem{Pal:2015mda}
S.~S. Pal and S.~Panda, {\it {Entanglement temperature with Gauss--Bonnet
  term}},  {\em Nucl. Phys.} {\bf B898} (2015) 401--414,
  [\href{http://arxiv.org/abs/1507.06488}{{\tt arXiv:1507.06488}}].

\bibitem{Matsueda:2015bba}
H.~Matsueda, {\it {Hessian potential for Fefferman-Graham metric}},
  \href{http://arxiv.org/abs/1508.06515}{{\tt arXiv:1508.06515}}.

\bibitem{Li:2015ola}
G.-Q. Li, J.-X. Mo, and X.-B. Xu, {\it {Entanglement temperature for black
  branes with hyperscaling violation}},
  \href{http://arxiv.org/abs/1509.05985}{{\tt arXiv:1509.05985}}.

\bibitem{Mansoori:2015sit}
S.~A.~H. Mansoori, B.~Mirza, M.~D. Darareh, and S.~Janbaz, {\it {Entanglement
  Thermodynamics of the Generalized Charged BTZ Black Hole}},
  \href{http://arxiv.org/abs/1512.00096}{{\tt arXiv:1512.00096}}.

\bibitem{Park:2015hcz}
C.~Park, {\it {Thermodynamic law from the entanglement entropy bound}},
  \href{http://arxiv.org/abs/1511.02288}{{\tt arXiv:1511.02288}}.

\bibitem{Tetradis:2009bk}
N.~Tetradis, {\it {The Temperature and entropy of CFT on time-dependent
  backgrounds}},  {\em JHEP} {\bf 03} (2010) 040,
  [\href{http://arxiv.org/abs/0905.2763}{{\tt arXiv:0905.2763}}].

\bibitem{deHaro:2000xn}
S.~de~Haro, S.~N. Solodukhin, and K.~Skenderis, {\it {Holographic
  reconstruction of space-time and renormalization in the AdS / CFT
  correspondence}},  {\em Commun. Math. Phys.} {\bf 217} (2001) 595--622,
  [\href{http://arxiv.org/abs/hep-th/0002230}{{\tt hep-th/0002230}}].

\bibitem{Skenderis:2000in}
K.~Skenderis, {\it {Asymptotically Anti-de Sitter space-times and their stress
  energy tensor}},  {\em Int. J. Mod. Phys.} {\bf A16} (2001) 740--749,
  [\href{http://arxiv.org/abs/hep-th/0010138}{{\tt hep-th/0010138}}].

\bibitem{Myers:1999psa}
R.~C. Myers, {\it {Stress tensors and Casimir energies in the AdS / CFT
  correspondence}},  {\em Phys. Rev.} {\bf D60} (1999) 046002,
  [\href{http://arxiv.org/abs/hep-th/9903203}{{\tt hep-th/9903203}}].

\bibitem{Hawking:1995fd}
S.~W. Hawking and G.~T. Horowitz, {\it {The Gravitational Hamiltonian, action,
  entropy and surface terms}},  {\em Class. Quant. Grav.} {\bf 13} (1996)
  1487--1498, [\href{http://arxiv.org/abs/gr-qc/9501014}{{\tt gr-qc/9501014}}].

\end{thebibliography}\endgroup
\end{document}